# Influence of Silicon Interlayers on Transition Layer Formation in Ti/Ni Multilayer Structures of Different Thicknesses

S.S. Sakhonenkov,[a] A.U. Gaisin,[a] A.S. Konashuk,[a] A.V. Bugaev,[a] R.S. Pleshkov,[b] V.N. Polkovnikov[b] and E.O. Filatova[a]

[a] *Institute of Physics, St-Petersburg State University, Ulyanovskaya Str. 1, Peterhof, St. Petersburg 198504, Russia*

[b] *Institute for Physics of Microstructure, Russian Academy of Sciences, Nizhny Novgorod 603087, Russia*

**Abstract**

This study presents a comprehensive investigation of chemical, structural, and magnetic properties of Ti/Ni multilayer systems with period thicknesses of 4 nm and 10 nm. Particular attention was paid to the characterization of the transition layers at Ni-Ti interfaces and the influence of thin silicon barrier layers on their formation. A combination of X-ray photoelectron spectroscopy (XPS), X-ray diffraction (XRD), X-ray reflectometry (XRR), and SQUID magnetometry was employed for analysis. Extended transition layers up to 1.2 nm in thickness were identified at the Ni-Ti interfaces, primarily composed of the intermetallic Ni$_3$Ti phase. The insertion of ultra-thin silicon buffer layers at the interfaces significantly suppressed the formation of intermetallic compounds, most likely due to the formation of titanium silicides. Additionally, it was observed that the use of Si layer on the sample surface leads to the formation of silicon oxide after exposure to the ambient environment, which acts as a passivation layer and inhibits oxidation of Ni and Ti layers within the topmost period of the multilayer structure.

## 1. Introduction

The combination of the physical properties of titanium and nickel makes them promising materials for use in multilayer structures applied in various fields. The absorption edge of titanium lies within the "water window", which is defined by the K-edge of oxygen (530 eV) and the K-edge of carbon (280 eV). Near this edge, titanium exhibits minimal X-ray absorption, making it a suitable candidate for use as a spacer in multilayer X-ray mirrors. When paired with Ni, a relatively large step in the real part of the refractive index occurs at the material interface [1]. As a result, this material combination demonstrates high theoretical reflectivity, which serves as the basis for its use in multilayer reflective components in optical schemes of X-ray microscopes [2–4], as well as in transmission phase shifters used in beamlines for synchrotron radiation [5]. From the perspective of neutron interaction, the Ti/Ni pair is the most suitable for use as a non-polarizing multilayer supermirror. This is because these materials exhibit the greatest difference in the real part of the neutron coherent scattering length density, i.e., they possess the highest neutron-optical contrast [6]. As a result, Ti/Ni multilayer coatings are used as optical elements for transporting, collimating, focusing, and monochromatizing neutron beams.

In the applications described above, it is crucial to maintain a high optical/neutron contrast between the layers; otherwise, the reflective performance of the multilayer mirrors will be significantly reduced. Several processes occurring at the interfaces between the layers can lead to a decrease in optical contrast. These include interdiffusion of atoms between adjacent layers, mixing during the deposition of films, and the formation of surface roughness, which is influenced by the crystallization of the layers. The resulting roughness, combined with the mixed layer, is referred to as the transition region. Numerous studies have shown that Ti/Ni mirrors exhibit extended asymmetric transition layers at the interfaces: approximately 0.7 nm for the Ti-on-Ni interface and about 1.2 nm for the Ni-on-Ti interface [7–10]. Depending on the deposition parameters, "roughness accumulation" can also occur, meaning that as new layers grow, their surface roughness will add to the roughness of the underlying layer [11,12].

Various interface engineering methods are employed to address the issue of transition layers: the use of thin buffer layers, doping layers with atoms of other elements, and ion-assisted deposition. For instance, the introduction of nitrogen, carbon, or oxygen into nickel, as well as vanadium or hydrogen into titanium, helps suppress interdiffusion, sharpens the interfaces, and stabilizes them [13]. A similar effect is achieved by doping the layers with atoms of boron ($^{11}$B) and carbon (C), as well as through ion assistance during magnetron sputtering deposition [10,14]. In the study by Stendahl et al. [15], the use of a $^{11}B_4C$ barrier layer reduced the thickness of the transition layer, although the resulting neutron-optical characteristics of the structure were unsatisfactory. It was shown that the insertion of $^{11}B_4C$ interlayers between relatively thick Ni and Ti layers, and the introduction of B and C into the structure in the case of thin layers, significantly improved the reflection properties. Furthermore, the study on the influence of thin silicon interlayers [16] demonstrated that the interlayer acts as a diffusion barrier at the Ni-on-Ti interface and as a smoothing layer at the Ti-on-Ni interface.

This work presents a comprehensive study of the chemical compounds formed in the transition layer between Ni and Ti, as well as an analysis of the influence of an inserted thin silicon buffer layers on the formation of these compounds. To obtain the necessary information, a set of complementary experimental methods was employed, including X-ray photoelectron spectroscopy (XPS), X-ray diffraction (XRD) and X-ray reflectometry (XRR). Considering the ferromagnetic nature of nickel, magnetometry measurements were also carried out to provide additional insight into the structural and magnetic properties of the multilayers. In the study conducted by Smertin and co-authors [16], the primary focus was on evaluating interface roughness and diffusion parameters, without providing any information on the chemical composition, as well as on assessing the impact of Si interlayers in a multilayer system with a single period thickness. A "period" is defined as the smallest repeating layer sequence. Since Ti/Ni structures with varying layer thicknesses are used for different applied tasks, this work focuses on comparing structures with two different period thicknesses: 4 nm and 10 nm. The results obtained significantly expand the understanding of the chemical structure of transition layers and their dependence on the introduction of Si barrier layers and the parameters of the multilayer system. This represents a valuable contribution to the development of new multilayer mirrors capable of providing improved performance in both X-ray and neutron optics, which is especially relevant for creating highly efficient optical elements in modern facilities. Moreover, insights into the formation of intermetallic phases in Ti/Ni systems are highly relevant for advancing their application in shape memory alloys [17,18].

## 2. Experimental

Samples in this study were synthesized via DC magnetron sputtering in an argon gas environment (purity 99.999%) at a pressure of ~0.1 Pa. Prior to synthesis, the residual gas pressure was maintained at $10^{-5}$ Pa, minimizing impurity incorporation (notably oxygen) into the structure. Four targets were used for sputtering: Ni, Ti and two Si, enabling deposition of a thin silicon film at both interfaces. Discharge current, magnetron voltage, and estimated target sputtering rates during synthesis are provided in Table 1. Detailed descriptions of the experimental setup are reported elsewhere [19,20]. All samples were prepared on smooth silicon substrates (root-mean-square roughness ~0.2 nm). The resulting multilayer systems are denoted as $[X/Y/Z]_n$, where X, Y and Z represent layer materials, and n is the number of periods. In the presented entry, the layers located closer to the substrate are indicated on the left, and those located closer to the surface are indicated on the right.

**Table 1.** Discharge currents I, magnetron voltages U and deposition rates for different targets.

| Target | I, mA | U, V | DR, nm / s |
|--------|-------|------|------------|
| Ni     | 0.6   | 256  | 0.1        |
| Ti     | 0.6   | 249  | 0.05       |
| Si     | 0.6   | 269  | 0.04       |

Chemical analysis of the structures was performed using X-ray photoelectron spectroscopy (XPS) on the ESCA module at the NanoPES station of the Kurchatov Synchrotron Radiation Source. The module is equipped with a PHOIBOS 150 hemispherical energy analyzer, providing an energy resolution of up to 2 meV. The source is an X-ray tube with an aluminum anode and a monochromator, which provides a monochromatic beam of X-ray photons with an energy of 1486.6 eV, a resolution of $10^{-4}$ and a beam spot on the sample of $1 \times 3$ mm$^2$. The angle between the incident X-ray beam and the analyzer axis is 55°. The pressure in the measuring chamber was maintained below $10^{-7}$ Pa. For surface cleaning and depth profiling, the module includes an ion gun with a gas inlet and differential pumping. As all samples in this study contained metal layers, energy calibration of the photoelectron spectra was based on the position of inflection point of a valence band. Background subtraction of inelastically scattered photoelectrons was performed using the universal Tougaard function [21,22]. In cases where peak intensities varied significantly for different samples, spectra were multiplied by appropriate scaling factors for ease of visual comparison.

Information on the crystalline structure and layer thicknesses was derived from the analysis of X-ray diffraction (XRD) and X-ray reflectometry (XRR) curves. Measurements were carried out on a Bruker D8 Discover laboratory diffractometer at the Resource Center for X-ray Diffraction Studies of Saint Petersburg State University. The instrument is equipped with an X-ray tube with a copper anode. A horizontal goniometer enabled θ-2θ geometry measurements, with sample positioning accuracy of 50 μm and an angular precision of 0.001°. The measured XRR reflection curves were fitted using the Multifitting software [23,24].

Magnetometry measurements were carried out using a superconducting quantum interference device magnetometer MPMS SQUID VSM manufactured by Quantum Design. The measurements were performed at the Centre for Diagnostics of Functional Materials for Medicine, Pharmacology and Nanoelectronics of Saint Petersburg State University. This system enables precise measurements of magnetic susceptibility over a wide range of temperatures and magnetic fields (±70 kOe). The sensitivity reaches $1 \cdot 10^{-8}$ emu at 0 Oe. The maximum sample size is 3 mm.

## 3. Results and Discussion

A total of four Ti/Ni multilayer structures were synthesized for this study. The nominal thicknesses of the individual layers were determined based on calibrated deposition rates and the substrate translation velocity over the targets. In the samples without silicon interlayers, the thicknesses of the Ni and Ti layers were designed to be equal within each period. To examine the influence of spatial scale on parameters of transition layers, two values of the period thickness were selected: 4 nm and 10 nm. The total thickness of the multilayer coatings was kept constant; therefore, the first system contains 105 periods, while the second one comprises 42 periods. To preserve the designed period thickness in the samples containing thin silicon interlayers, the nominal thickness of the Ti layer was decreased by an amount equal to the total nominal thickness of the Si layers incorporated within each period. The reduction specifically of the titanium layer is attributed to the fact that Ti and Si exhibit closer neutron-optical characteristics, than Ni and Si.

Table 2. Nominal parameters of synthesized samples: d - period, $t_i$ - thickness of layer i.

| № | Structure | d, nm | $t_{Ti}$, nm | $t_{Ni}$, nm | $t_{Si}$, nm |
|---|---|---|---|---|---|
| 1 | [Ti/Ni] | 4 | 2 | 2 | |
| 2 | [Ti/Si/Ni/Si] | 4 | 1.4 | 2 | 0.3 |
| 3 | [Ti/Ni] | 10 | 5 | 5 | |
| 4 | [Ti/Ni/Si/Ni] | 10 | 4.2 | 5 | 0.4 |

### 3.1 X-ray Photoelectron spectroscopy

The interaction between Ni and Ti can result in the formation of stable intermetallic compounds. First-principles calculations performed using the Vienna *Ab Initio* Simulation Package (VASP) [25] show that $Ni_xTi$ intermetallics exhibit negative formation energies, with $Ni_3Ti$ demonstrating the lowest value among the considered phases (Figure 1). This thermodynamic result is further supported by kinetic studies [26,27], which reveal that $Ni_3Ti$ also possesses the lowest diffusion activation energy, explaining its preferential nucleation during the initial stages of the interfacial reaction in Ti/Ni multilayers. The subsequent formation of $NiTi_2$ and $NiTi$ phases occurs only after near-interface Ni depletion, following a sequential phase evolution mechanism.

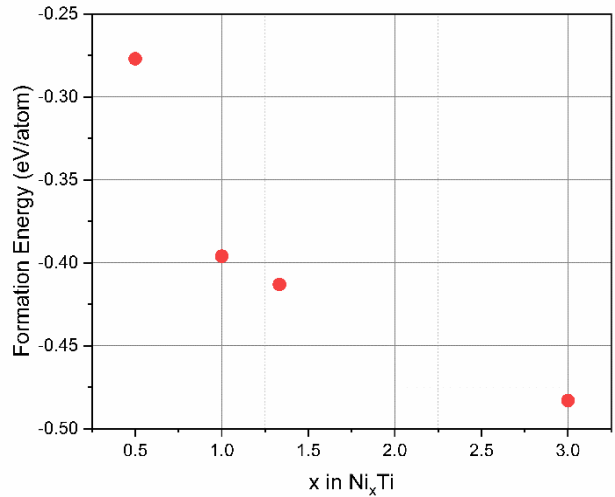

**Figure 1.** Formation energy for various Ni$_x$Ti intermetallics [25].

The stoichiometry of Ni$_x$Ti intermetallic compounds can be evaluated using X-ray photoelectron spectroscopy (XPS) by analyzing the binding energy shifts of Ni 2p and Ti 2p core-level peaks relative to their positions in pure metals. According to [28], a higher Ni content in the intermetallic phase results in a smaller Ni 2p shift but a more pronounced Ti 2p shift toward higher binding energies. Conversely, Ti-rich compounds exhibit the opposite trend. Similar trends in peak shifts of Ni$_x$Ti intermetallics were shown in the articles [14,29,30]. An additional indicator of intermetallic formation is the increased energy separation between the main Ni 2p peak and its satellite: from 5.8 - 6 eV in pure Ni to ~7 eV in NiTi intermetallic [29,31,32]. The satellite peak arises due to the creation of an additional 3d-hole state during photoemission [31,33].

To obtain photoelectron spectra from elemental Ni, Ti, and Ni$_x$Ti intermetallic under identical experimental conditions, samples consisting of thin film deposited onto silicon substrates were synthesized. The intermetallic film was synthesized as a Ti/Ni multilayer structure with ultra-thin nominal layers (approximately 0.3 nm thick) and 200 periods. Using magnetron sputtering deposition, the impinging atoms possessed sufficient energy to promote interfacial mixing, resulting in the formation of an extended intermixed layer. The nominal layer thicknesses were specifically chosen to be less the characteristic thickness of this intermixed region (from 0.7 nm to 1.2 nm, as mentioned in the introduction), ultimately yielding a homogeneous intermetallic film. Considering that the nominal thicknesses of Ni and Ti layers were equal, and that the atomic concentration of Ni ($9.14 \cdot 10^{22}$ cm$^{-3}$) is approximately 1.6 times higher than that of Ti ($5.66 \cdot 10^{22}$ cm$^{-3}$) [34], it can be expected that the resulting film represents a Ni$_x$Ti intermetallic with a relatively higher nickel content.

Figure 2a shows the Ni 2p$_{3/2}$ photoelectron lines obtained for the Ni and Ni$_x$Ti films. The film surfaces were cleaned using argon ion sputtering. The ion beam was directed at the sample at a grazing incidence angle of 30°, with an energy of 750 eV and total sputtering duration of 10 minutes. Analysis of spectrum after ion sputtering reveal the energy position of the elemental nickel peak at 852.75 eV, with its satellite peak shifted by 5.75 eV toward higher binding energies. The main peak has a tail on the higher binding energy side, indicating the presence of residual nickel oxide/hydroxide remaining after sputtering. The surface oxidation of the samples occurred due to their exposure to the ambient environment during transportation and sample preparation before measurements. The main line of the Ni$_x$Ti intermetallic is shifted by 0.15 eV toward higher binding energies relative to the main elemental nickel line. The energy separation from its satellite increases to 6.2 eV.

Figures 2b and 2c present the Ti 2p photoelectron spectra obtained for Ti and Ni$_x$Ti films, before and after sputtering, respectively. In Figure 2c and subsequent images, notations such as "x3.5" indicate the factor by which the corresponding spectrum has been multiplied. It can be seen that before sputtering the film surfaces were heavily oxidized, which is evidenced by an intense peak at a binding energy of 459.3 eV. This peak can be attributed to a signal from TiO$_2$ [35,36]. Prior to sputtering, the energy positions of the titanium and intermetallic peaks could be roughly estimated at 454.5 eV and 454.8 eV, respectively. In the post-sputtering spectra, the intermetallic peak shifts slightly and is observed at 455.0 eV. It should be noted that the peaks in the post-sputtering spectra are relatively broad. Combined with the fact that a significant O 1s line signal remains (not shown), one can conclude that the spectra contain substantial contributions from titanium suboxides. According to the literature [37–40], the Ti 2p$_{3/2}$ peaks of these compounds appear at binding energies of 454.7 - 457.3 eV, with their corresponding Ti 2p$_{1/2}$ components separated by 5.2-5.6 eV. Consequently, precise determination of the metal/intermetallic peak positions is complicated by strong overlap with suboxide peaks, which also explains the observed peak maximum shift after sputtering. Nevertheless, a significant shift of the intermetallic peak toward higher binding energies relative to the metal peak is clearly visible. Considering that the peak shift in the Ni 2p line of the intermetallic is notably smaller than that in the Ti 2p line, one can conclude that the spectrum corresponds to an intermetallic with a relatively higher nickel content.

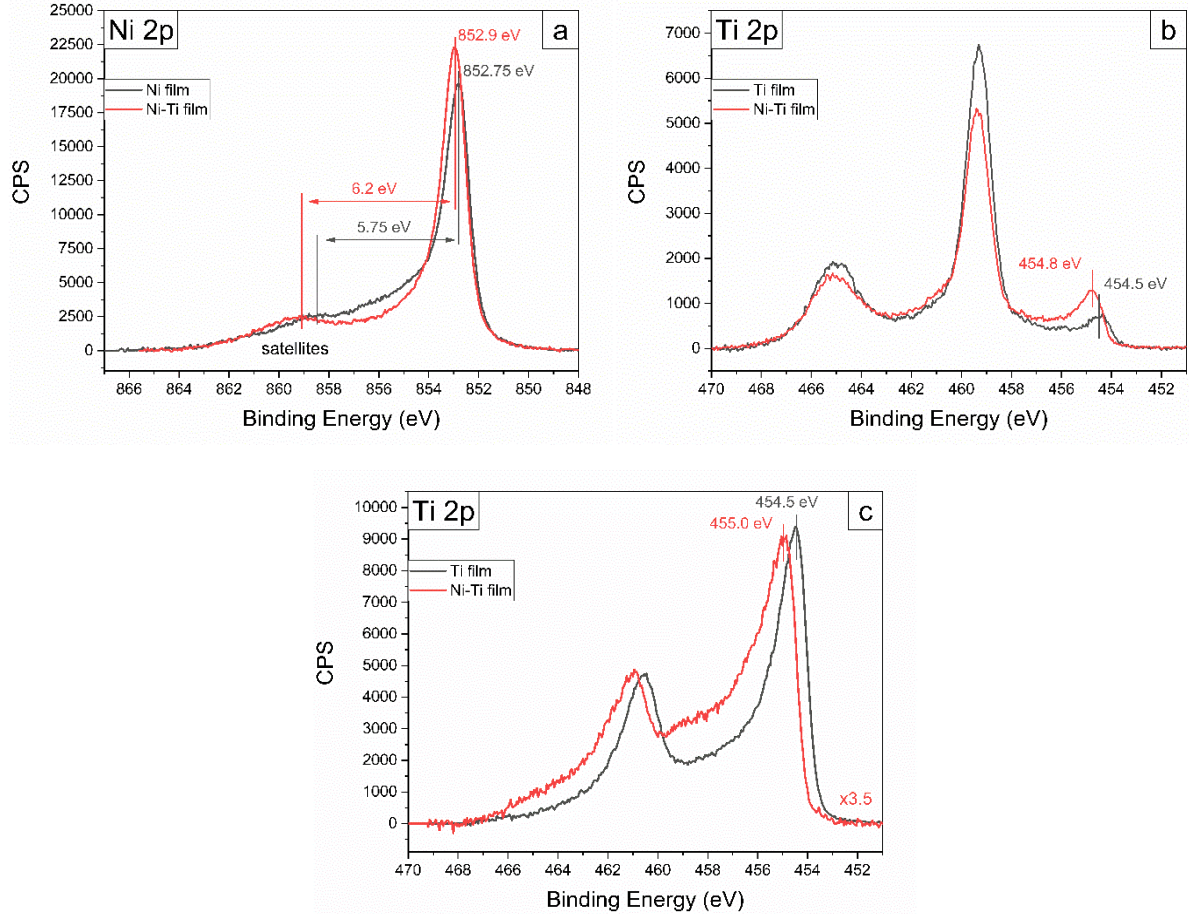

**Figure 2.** (a) Ni 2p$_{3/2}$ and (b, c) Ti 2p photoelectron spectra of Ni, Ti and Ni$_x$Ti films. Spectra in (a) and (c) were obtained after cleaning the film surfaces by argon ion sputtering.

An additional confirmation of this conclusion can be derived by analyzing the intensity ratios of the Ni 2p$_{3/2}$ and Ti 2p$_{3/2}$ peaks related to the Ni$_x$Ti intermetallic. The stoichiometry of thin films is determined using the following relation:

$$\frac{n_a}{n_b} = \frac{I_a}{RSF_a} \bigg/ \frac{I_b}{RSF_b}$$

where $n_i$ – is the atomic concentration of element $i$, $I_i$ – is the photoelectron line intensity for atoms $i$, $RSF$ – relative sensitivity factor, incorporating partial ionization cross section for the orbital concerned $\sigma$, an angular distribution term $\varphi$, the efficiency of detection of the spectrometer $T$ and inelastic mean free path $\lambda$ [41]. Photoionization cross-section and asymmetry parameters are sourced from [42], while IMFP is estimated using the TPP-2M formula [43]. The energy analyzer efficiency is experimentally determined and approximated as a power function dependent on kinetic energy, $KE^{-n}$, where n ranges from 0.4 до 1.1 [44]. Since the efficiency of energy analyzer in the experimental setup used in these studies was not precisely evaluated, calculations were performed multiple times with varying n values (minimum 1.1, maximum 0.4, and average 0.7). The results indicate that the as-deposited sample is a $Ni_xTi$ film with $x$ ranging from approximately 1.7 to 2.4, while the sputtered sample yields $x$ between 3.0 and 4.2. It is evident that in both cases an intermetallic compound with a high nickel content is formed. The noticeable difference in the x value before/after sputtering is most likely due to the large overlap of the intermetallic peaks with the suboxide peaks, which makes it difficult to isolate them accurately.

Thus, the spectra of pure Ni, Ti, and a $Ni_xTi$ intermetallic with a high relative nickel content were measured as references under identical experimental conditions.

Figures 3a and 3b compare the Ni $2p_{3/2}$ and Ti 2p spectra of the reference films and the $[Ti/Ni]_{105}$ multilayer structure with a 4 nm period. The main narrow intense peak in the Ni $2p_{3/2}$ line is broadened compared to the reference peaks, fully encompassing them. The Ti $2p_{3/2}$ peak for the multilayer lies between the reference peaks. Here, the reference peak widths are not considered in the analysis due to significant suboxide contributions. Nevertheless, the observed pattern suggests that Ni and Ti in the multilayer have interacted, forming a $Ni_xTi$ intermetallic. Both elemental Ni and Ti remain in the system. Estimating the transition layer thickness relative to the system period could involve decomposition the spectra and comparing metal/intermetallic peak intensities. However, due to the close proximity of these peaks and their overlap with suboxides, such decomposition would be unstable and potentially misleading. Notably, the multilayer spectra show intense components from nickel oxide/hydroxide (main component at 855.7 eV and satellite at 861.5 eV [45,46]) and $TiO_2$, indicating that the top Ni layer does not prevent oxidation of the underlying Ti layer. Notable oxidation of titanium beneath the nickel layer can be attributed to the fact that titanium has a higher tendency to oxidize compared to nickel, due to the more negative standard Gibbs free energy of formation for titanium oxide. Thus, at least the layers from the first period of the structure have undergone partial oxidation.

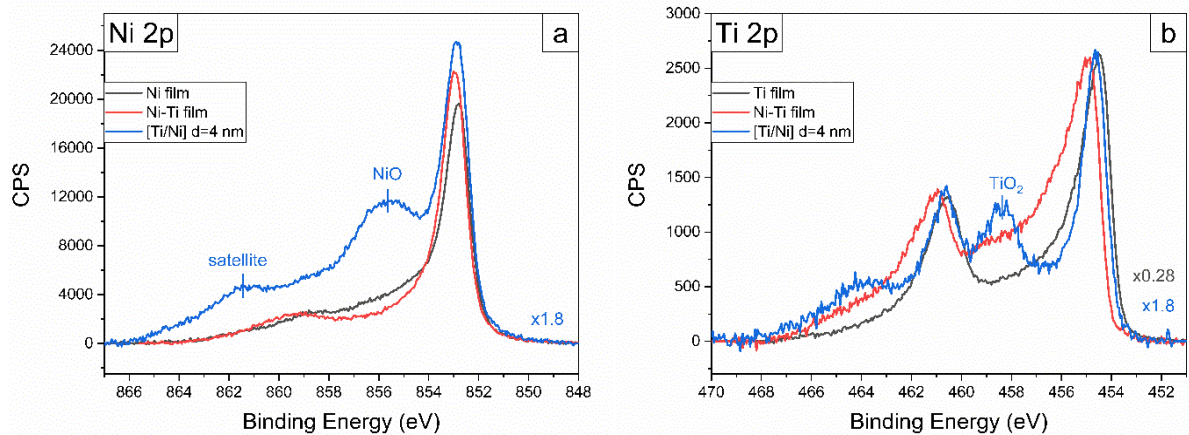

**Figure 3.** Photoelectron spectra of (a) Ni 2p$_{3/2}$ and (b) Ti 2p obtained for Ni, Ni$_x$Ti films and [Ti/Ni]$_{105}$ multilayer structure with a 4 nm period.

Similar spectra were obtained for the [Ti/Ni]$_{42}$ system with a 10 nm period. The Ni 2p$_{3/2}$ spectrum closely matches that of the previously discussed multilayer structure, differing only in the slightly lower intensity of the NiO component relative to Ni. The Ti 2p spectrum has very low intensity and high noise due to suppression by the overlying nickel, yet its shape nearly fully aligns with the d=4 nm sample spectrum. Thus, Ni$_x$Ti formation is evident, and even under a 5 nm nominal Ni layer, titanium layer still oxidizes.

Figures 4a and 4b show photoelectron spectra of Ni 2p$_{3/2}$ и Ti 2p obtained for multilayer structures [Ti/Ni]$_{105}$ and [Ti/Si/Ni/Si]$_{105}$ with 4 nm period. Silicon buffer layers (0.3 nm nominal thickness) were applied at each interface. In the system with silicon buffer layers, Ti 2p peaks shift slightly to lower binding energies, and the Ni peak's width decreases. The Ni 2p peak shape closely matches the main peak of the structure without buffer layer on the lower binding energy side, where elemental Ni contributes. This suggests reduced Ni$_x$Ti content. Taking into account that the Si layers are very thin, it probably reacted completely with Ni and/or Ti. According to VASP calculations [25], Ti-Si compounds are energetically favored: -0.808 eV/atom for Ti$_5$Si$_4$ and -0.524 eV/atom for Ni$_2$Si. Literature data on Ni and Ti silicides reveal that Ni silicide peaks in Ni 2p shift to higher binding energies (up to 2 eV for Si-rich phases), while Ti silicide peaks in Ti 2p shift to lower energies (up to -0.5 eV for TiSi$_2$) [47–49]. Based on the above analysis, the most likely scenario is the interaction of Si with Ti and formation of a titanium silicide. Notably, in the structure with Si interlayers, contributions from Ti and Ni suboxides are nearly absent. The Si 2s spectrum from the multilayer sample reveals two components (Figure 4c). The component at 154.65 эВ can be attributed to SiO$_x$, where x is approximately equal to 2 [50,51]. In the literature, an analysis of Si 2p spectra is usually given. In Si 2s, the relative energy positions of the peaks of silicon and its oxides approximately coincide. Surface SiO$_x$ acts as a passivation layer in the Ti/Ni structure, preventing Ni and Ti oxide formation. The second component, at 151.2 eV, aligns closely with elemental Si [52,53]. It is noteworthy that titanium silicide peaks are also proximate in energy position to elemental Si [54,55]. In the light of the above analysis, this peak may fully correspond to titanium silicide. However, without a reference spectrum of elemental Si measured under identical conditions, conclusively identifying the second component remains challenging. In this study, the measurements of spectra for elemental Si were not carried out.

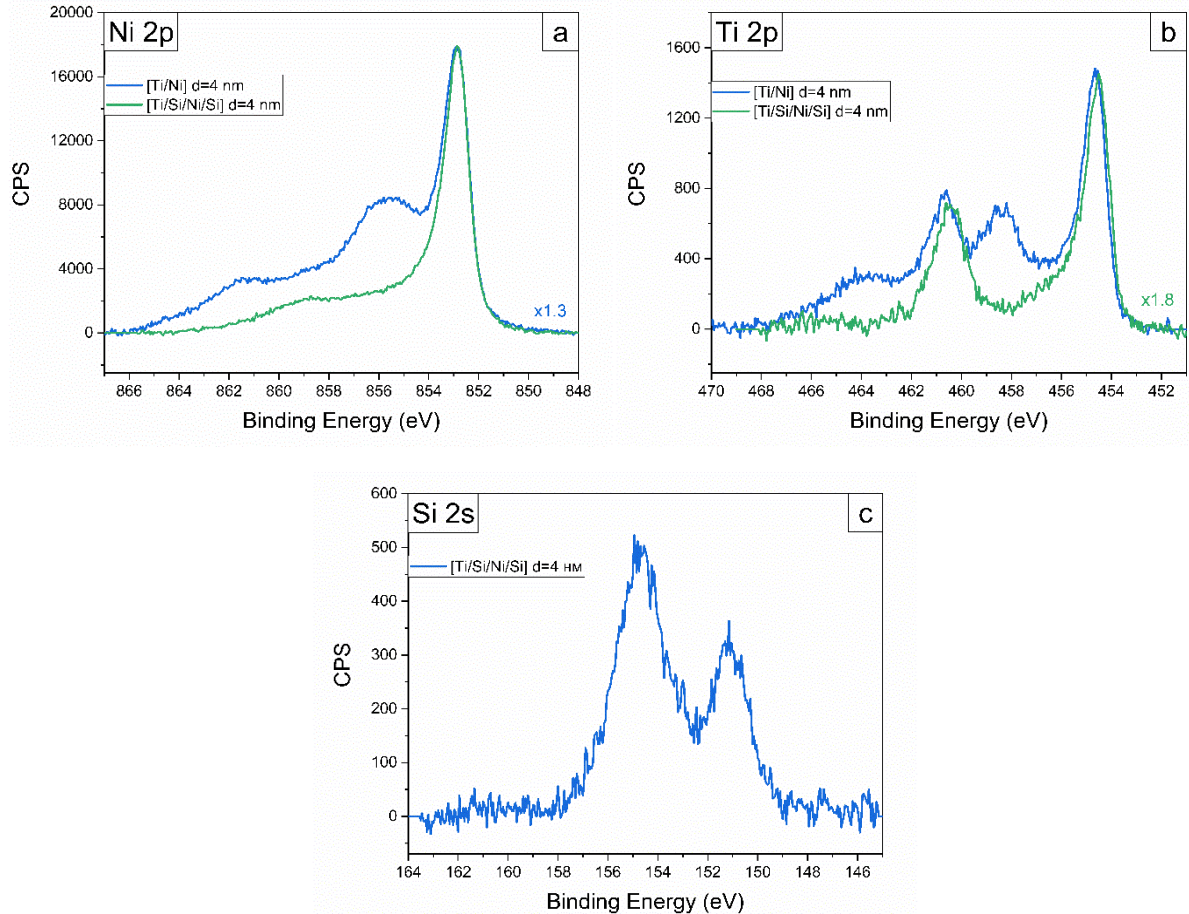

**Figure 4.** Photoelectron spectra of (a) Ni 2p$_{3/2}$, (b) Ti 2p and (c) Si 2s, obtained for multilayer structures [Ti/Ni]$_{105}$ and [Ti/Si/Ni/Si]$_{105}$ with a 4 nm period.

In the system with period thickness of 10 nm the nominal thickness of silicon layers is equal to 0.4 nm. The Ni 2p, Ti 2p, and Si 2s spectra exhibit trends identical to those observed previously: absence of titanium/nickel oxide contributions, presence of silicon oxide at the surface, a slight shift of Ti 2p to lower binding energies, and a reduced full width at half maximum of the main Ni 2p$_{3/2}$ peak in the interlayer system. Thus, this system also shows reduced Ni-Ti intermetallic formation due to titanium silicide formation.

### 3.2 X-ray Diffraction

The X-ray diffraction patterns of the [Ti/Ni]$_{42}$ and [Ti/Si/Ni/Si]$_{42}$ multilayer structures with d = 10 nm are presented in Figure 5a. Both curves consist of three clearly distinguishable peaks, as well as a less pronounced feature around the angle of 34.5°. The intense peak at 2θ = 38.2° and the low intensity peak at 34.5° can be assigned to the (002) plane and (010) plane, respectively, of hexagonal close-packed (α-Ti) titanium. Deposited titanium films and layers in multilayers structures of thickness less than 100 nm are usually highly textured, exhibiting a single dominant peak [56–58]. For the Si/[Ti/Si/Ni/Si]$_{42}$ multilayer, the main intensive peak at 2θ = 44.5° was assigned to the (111) plane of face-centered cubic Ni [59]. In contrast, the position of the main peak of the Si/[Ti/Ni]$_{42}$ structure is shifted to 2θ = 44.2°. To determine the nature of the shift and correctly identify the peak's origin, a decomposition of the diffractograms was carried out (Figure 5b, c). It is noteworthy that fitting the main peak with a single component is infeasible due to its slight asymmetry. Decomposition analysis revealed that the main peak in the patterns consists of two distinct components: one corresponding to the Ni (111) plane and the other to the (004) plane of hexagonal Ni$_3$Ti. Previous studies of Ti/Ni multilayer mirrors with different layer thickness [60–62] have also reported the possible formation of a hexagonal Ni$_3$Ti phase, based on X-ray diffraction and selected area electron diffraction data. The presence of this peak was also observed in XRD patterns of high-temperature annealed Ti/Ni multilayer periodic nanostructures [29,63,64]. From the presented decompositions it is evident that in the structure with thin Si layers the Ni$_3$Ti component is smaller.

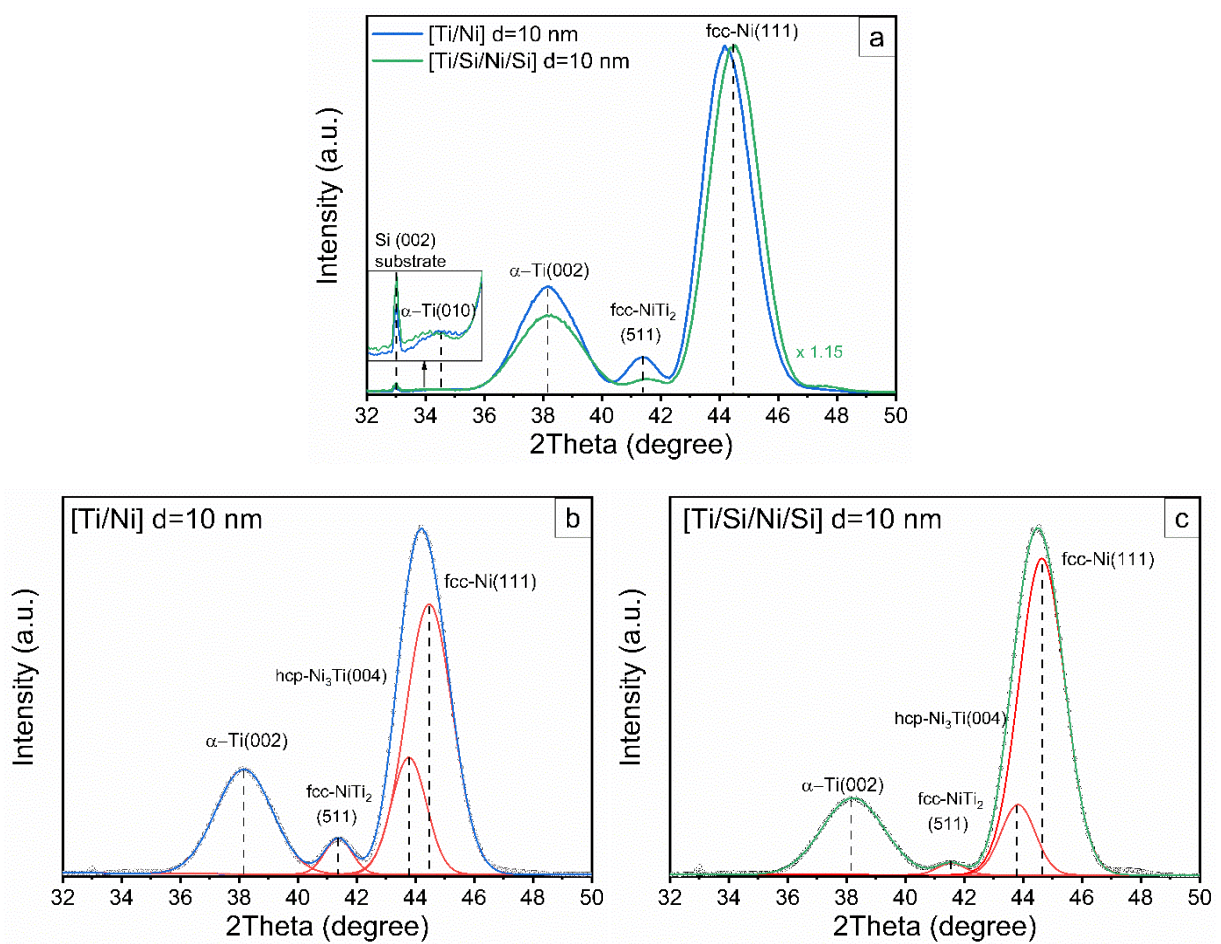

**Figure 5.** XRD patterns of [Ti/Ni]$_{42}$ and [Ti/Si/Ni/Si]$_{42}$ multilayer structures with d = 10 nm (a) and their decomposition into components (b, c).

The peak at $2\theta \approx 41.3° - 41.5°$ was observed in as-deposited Ti/Ni multilayer structures in previous studies [16,58], though its origin remained unexplained. An analysis of the literature shows that it is quite reasonable to associate this peak with reflection from (511) plane of the face-centered cubic (fcc) NiTi$_2$ phase (PDF № 00-018-0898) [65]. This is supported by studies on annealed multilayer structures, where the NiTi$_2$ phase content prevails over Ni$_3$Ti with increasing titanium-to-nickel thickness ratio [63]. Similar to the Ni$_3$Ti phase, the inserting of a Si barrier layer reduces the formation of fcc-NiTi$_2$ crystal structure.

**Table 3.** Results of X-ray phase analysis processing. FWHM – full width at half maximum, a – interplanar spacing, l$_{coh}$ – coherence length

| Reflection | Parameter | Structure d = 10 nm | |
|---|---|---|---|
| | | [Ti/Ni]$_{42}$ | [Ti/Si/Ni/Si]$_{42}$ |
| Ni (111) | 2θ, ° | 44.47 | 44.64 |
| | FWHM, ° | 1.87 | 1.84 |
| | a, Å | 2.04 | 2.03 |
| | L$_{coh}$, nm | 4.49 | 4.55 |
| Ti(002) | 2θ, ° | 38.17 | 38.22 |
| | FWHM, ° | 2.27 | 2.48 |
| | a, Å | 2.36 | 2.35 |
| | L$_{coh}$, nm | 3.62 | 3.31 |

Based on the decomposition results, the crystal structure of the main layers can be analyzed (Table 3). The interplanar spacing (002) of pure titanium and (111) nickel of two multilayer mirrors is identical within the experimental error limits. Using the Scherrer formula [66], we estimated the coherent scattering length (associated with crystallite size) from the full width at half maximum (FWHM) of the diffraction peaks. In the system with thin Si interlayers, the coherence length in the Ni layers remains nearly unchanged, whereas in the Ti layers it decreases, primarily due to the reduction in the nominal thickness of the titanium.

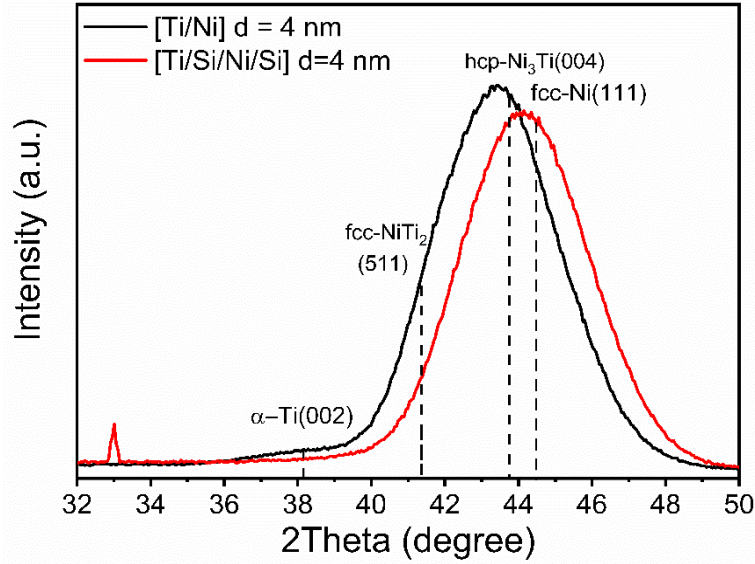

**Figure 6.** XRD pattern of [Ti/Ni]$_{105}$ and [Ti/Si/Ni/Si]$_{105}$ multilayer structure with d = 4 nm.

X-ray diffraction patterns of [Ti/Ni]$_{105}$ and [Ti/Si/Ni/Si]$_{105}$ multilayer structures with d = 4 nm are presented in Figure 6. The diffraction pattern of the mirrors is mainly characterized by a single broad peak. In the structure without barrier layers, this peak is shifted toward smaller angles of 2θ, likely due to contributions from various nickel-titanium intermetallic phases with different stoichiometries. The insertion of Si barrier layers shifts the diffraction peak towards the position of pure nickel (111), which indicates a decrease in the formation of intermetallic compounds. It is worth noting that in the [Ti/Ni]$_{105}$ structure, a weak peak corresponding to hexagonal Ti (002) plane is observable, while in the [Ti/Si/Ni/Si]$_{105}$ structure the peak of crystalline titanium is not detected.

In light of the above analysis, we assume that buffer silicon layers inhibit the formation of the Ni$_3$Ti and NiTi$_2$ crystalline phases.

### 3.3 Magnetometry

The results of XPS and XRD studies demonstrate the effect of silicon barrier layers on the composition of the transition region in the Ti/Ni multilayer mirrors. It is well known that the ferromagnetic properties of nickel impose a limitation on the use of Ti/Ni reflective coatings in polarized neutron optics, since in-plane magnetization leads to beam depolarization. Therefore, it is of particular interest to examine how the introduction of silicon barrier layers at both interfaces influences the overall magnetization of the multilayer structure.

For this purpose, SQUID magnetometry measurements were performed at room temperature on multilayer nanostructures with a period thickness of 10 nm. Figure 7 shows the magnetic field dependence of the volume magnetization ($M_{vol}$), measured in the plane of the [Ti/Ni]$_{42}$ and [Ti/Si/Ni/Si]$_{42}$ multilayer structures. The volume magnetization was calculated by normalizing the measured magnetic moment to the total nominal Ni thickness and the sample area. As follows from the obtained data, the remanent volume magnetization ($M_{Rvol}$) in the [Ti/Ni]$_{42}$ multilayer is lower than that in the system containing silicon barriers, indicating a reduced volume of ferromagnetic Ni or magnetic compounds of nickel with titanium in the sample without Si layers. The coercive field ($H_c$) for the multilayers with and without silicon interlayers was determined to be 30.4 Oe and 28.5 Oe, respectively. Previous studies [64,67] attributed a decrease in coercivity with increasing annealing temperature to a reduction in the thickness of pure Ni layers, caused by the formation of intermetallic compounds at the interfaces. Ferromagnetic ordering in ultrathin Ni layers is highly sensitive to thickness, and for Ni films thinner than 3 nm, magnetism is typically suppressed [68]. Therefore, the observed change in coercivity may be attributed to the reduced thickness of pure Ni in the [Ti/Ni]$_{42}$ multilayer compared to that in the [Ti/Si/Ni/Si]$_{42}$ structure.

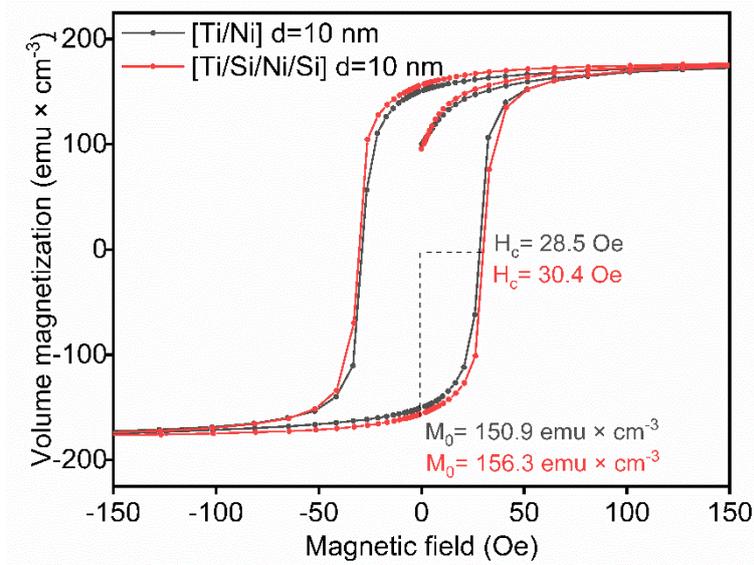

**Figure 7.** Volume magnetization as a function of the in-plane applied magnetic field for the [Ti/Ni]$_{42}$ and [Ti/Si/Ni/Si]$_{42}$ multilayers with period d=10 nm.

### 3.4 X-ray Reflectometry

To evaluate the thicknesses of both transitional and base layers in the studied structures, a fitting procedure of experimental X-ray reflectivity (XRR) curves was employed. Theoretical models were constructed based on information obtained from XPS and XRD analysis. In all models, a surface contamination layer of carbon was included. For the systems without silicon barrier layers, the top layers were modeled in the following sequence (from substrate to surface): $Ni_xTi/Ti/Ni_xTi/TiO_2/Ni/NiO$. The alternating multilayer part was described using the model $[Ni_xTi/Ti/Ni_xTi/Ni]_n$, where n was set to 104 and 41 for the samples with nominal period thicknesses of 4 nm and 10 nm, respectively. In the systems with a silicon barrier layer, a $SiO_2$ layer was added below the surface carbon layer. Below that, the multilayer stack was modeled as $[Ni_xTi/TiSi_y/Ti/TiSi_y/Ni_xTi/Ni]_n$, with *n* = 105 and 42. In all cases, a silicon substrate with a fixed density of 2.2 g/cm³ was assumed. During the fitting procedure, the densities of the base layers, oxide layers, and carbon contamination layer were fixed to tabulated values and not varied: Ti - 4.5 g/cm³, Ni - 8.9 g/cm³, NiO - 6.67 g/cm³, $TiO_2$ - 4.26 g/cm³, and C - 2.2 g/cm³ [69]. The densities of $TiSi_y$ and $Ni_xTi$ were allowed to vary within the range of tabulated values: 4.02 g/cm³ ($TiSi_2$) to 4.37 g/cm³ ($Ti_5Si_3$), and 6.45 g/cm³ (NiTi) to 8.17 g/cm³ ($Ni_3Ti$) [25].

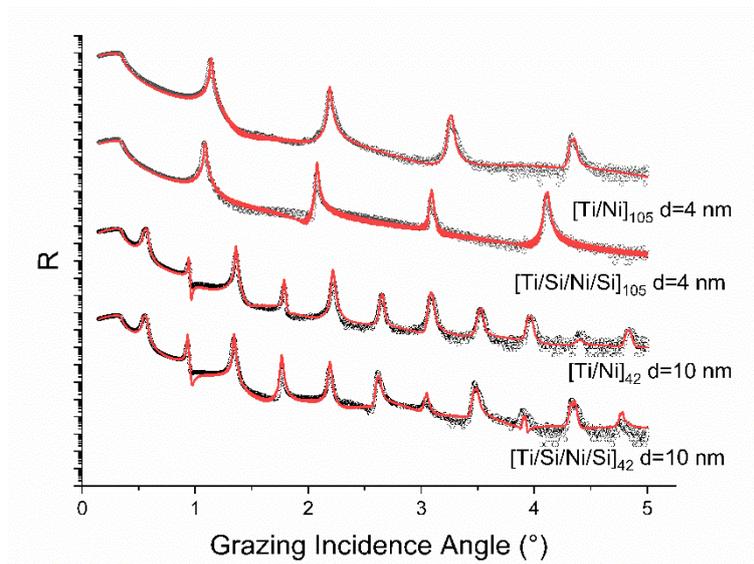

**Figure 8.** Experimental reflectivity curves (dots) and corresponding fits (red lines) for [Ti/Ni] multilayer structures with and without Si barrier layers. The d values on the graph are nominal. The curves are vertically offset for clarity and ease of visual comparison.

Figure 8 presents the experimental and fitted reflectivity curves for all studied samples. Table 4 summarizes the fitting parameters obtained for the alternating layer models. The results show that, in the systems without Si interlayers, an extended intermetallic compound forms between Ni and Ti. In structures with different period thicknesses, the transition layers are asymmetric, with comparable thicknesses of approximately 1.2 nm at the Ni-on-Ti interface and about 0.6 nm at the Ti-on-Ni interface. The insertion of silicon buffer layer leads to the formation of a titanium silicide layers of equal thickness at both interfaces. Moreover, in the system with the thicker period, the silicide layer is more extended (0.6 nm vs. ~0.3 nm). The nominal thickness of the silicon layers also differed: 0.3 nm for the sample with thinner layers and 0.4 nm for the other. In the Si-containing samples, the Ni layer thickness increased, whereas the thickness of the intermetallic and Ti layers decreased. All of these trends were also observed in the preceding XPS, XRD, and magnetometry analyses.

Table 4. Summary of fitting parameters for the alternating parts of the multilayer samples: $t$ - layer thickness, $\sigma$ - interfacial roughness, $\rho$ - layer density, $d$ - period thickness.

|  | composition | $Ni_xTi$ | $TiSi_y$ | Ti | $TiSi_y$ | $Ni_xTi$ | Ni |
|---|---|---|---|---|---|---|---|
| [Ti/Ni]$_{105}$ d=4.09 nm | t, nm | 0.50 |  | 1.49 |  | 1.15 | 0.95 |
|  | σ, nm | 0.28 |  | 0.44 |  | 0.37 | 0.26 |
|  | ρ, g/cm$^3$ | 7.8 |  |  |  | 8.1 |  |
| [Ti/Si/Ni/Si]$_{105}$ d=4.32 nm | t, nm | 0.25 | 0.27 | 1.01 | 0.26 | 0.77 | 1.76 |
|  | σ, nm | 0.26 | 0.20 | 0.04 | 0.20 | 0.19 | 0.06 |
|  | ρ, g/cm$^3$ | 8.10 | 4.30 |  | 4.30 | 7.90 |  |
| [Ti/Ni]$_{42}$ d=10.08 nm | t, nm | 0.66 |  | 4.85 |  | 1.21 | 3.36 |
|  | σ, nm | 0.27 |  | 0.39 |  | 0.29 | 0.24 |
|  | ρ, g/cm$^3$ | 7.80 |  |  |  | 8.03 |  |
| [Ti/Si/Ni/Si]$_{42}$ d=10.25 nm | t, nm | 0.40 | 0.60 | 4.01 | 0.60 | 0.56 | 4.05 |
|  | σ, nm | 0.28 | 0.30 | 0.12 | 0.40 | 0.40 | 0.58 |
|  | ρ, g/cm$^3$ | 7.00 | 4.5 |  | 4.30 | 7.02 |  |

4. **Conclusion**

A comprehensive chemical and structural investigation of transition regions in Ti/Ni multilayer structures with different bilayer thicknesses was carried out, along with an analysis of the effect of inserting thin silicon buffer layers. The study employed complementary techniques including X-ray photoelectron spectroscopy (XPS), X-ray diffraction (XRD), X-ray reflectometry (XRR), and magnetometry.

The results revealed the formation of an extended transition layer between Ni and Ti, with asymmetry in its thickness: the Ni-on-Ti interface exhibited a thicker transition zone (~1.2 nm) compared to the Ti-on-Ni interface (~0.6 nm). This asymmetry was observed consistently in both systems with nominal bilayer periods of approximately 4 nm and 10 nm. The transition region predominantly consists of the intermetallic compound $Ni_3Ti$. In the 10 nm-period structure, crystallization of the Ni and Ti layers occurs, accompanied by the formation of crystalline $Ni_3Ti$ and a smaller amount of $NiTi_2$. In contrast, in the 4 nm-period system, crystallization of the titanium layer is largely suppressed.

The insertion of thin silicon buffer layers at both interfaces significantly suppressed the formation of intermetallic phases. It is likely that silicon does not remain in its elemental form, but instead reacts with titanium to form titanium silicides. In the 4 nm-period system with silicon interlayers, crystallization of the titanium layers was completely inhibited. It is also worth noting that in the topmost period of the multilayer structure, oxidation occurs in both the nickel and titanium layers with formation of NiO and TiO$_2$. Applying silicon onto the nickel surface completely prevents oxidation of these layers by forming a passivating silicon oxide layer.

These findings demonstrate the effectiveness of silicon interlayers in stabilizing the structure and improving its potential performance in applications involving X-ray and neutron optics.

**Funding sources**

Russian Science Foundation Grant No. 24-72-10107.

**Acknowledgements**

This work was supported by the Russian Science Foundation (RSF), grant No. 24-72-10107. The authors express their sincere gratitude to the staff of the "Centre for X-ray Diffraction Studies" and the "Centre for Diagnostics of Functional Materials for Medicine, Pharmacology and Nanoelectronics" at the Research Park of Saint Petersburg State University for assistance with XRD, XRR, and magnetometry measurements. The authors also gratefully acknowledge the National Research Center "Kurchatov Institute" for providing access to the ESCA laboratory equipment and for their valuable technical support.